\documentclass[%
superscriptaddress,
 amsmath,amssymb,
 aps,
prb,
twocolumn,
floatfix,
]{revtex4-2}
\usepackage{mhchem}
\usepackage{gensymb}
\usepackage{booktabs}
\usepackage{graphicx}
\usepackage{dcolumn}
\usepackage{bm}
\usepackage[dvipsnames]{xcolor}
\usepackage{float}

\begin{document}

\preprint{APS/123-QED}

\title{Fermi surface mapping and the nature of charge density wave order in the kagome superconductor CsV$_3$Sb$_5$}%

\author{Brenden R. Ortiz$^{\dagger}$}
 \email{ortiz.brendenr@gmail.com}
 \affiliation{Materials Department and California Nanosystems Institute, University of California Santa Barbara, Santa Barbara, CA, 93106, United States}%
 
\author{Samuel M. L. Teicher}
 \thanks{These authors contributed equally}
 \affiliation{Materials Department and California Nanosystems Institute, University of California Santa Barbara, Santa Barbara, CA, 93106, United States}%

\author{Linus Kautzsch}
 \affiliation{Materials Department and California Nanosystems Institute, University of California Santa Barbara, Santa Barbara, CA, 93106, United States}%

\author{Paul M. Sarte}
 \affiliation{Materials Department and California Nanosystems Institute, University of California Santa Barbara, Santa Barbara, CA, 93106, United States}%
 
 \author{Noah Ratcliff}
 \affiliation{Materials Department and California Nanosystems Institute, University of California Santa Barbara, Santa Barbara, CA, 93106, United States}%
 
 \author{John Harter}
 \affiliation{Materials Department and California Nanosystems Institute, University of California Santa Barbara, Santa Barbara, CA, 93106, United States}%

\author{Jacob P. C. Ruff}
 \affiliation{CHESS, Cornell University, Ithaca, NY, 14853, United States}%
 
\author{Ram Seshadri}
 \affiliation{Materials Department and California Nanosystems Institute, University of California Santa Barbara, Santa Barbara, CA, 93106, United States}%
 
\author{Stephen D. Wilson}
 \email{stephendwilson@ucsb.edu}
 \affiliation{Materials Department and California Nanosystems Institute, University of California Santa Barbara, Santa Barbara, CA, 93106, United States}%

\date{\today}

\begin{abstract}
The recently discovered family of \textit{A}V$_3$Sb$_5$ (\textit{A}: K, Rb Cs) kagome metals possess a unique combination of nontrivial band topology, superconducting ground states, and signatures of electron correlations manifest via competing charge density wave order. Little is understood regarding the nature of the charge density wave (CDW) instability inherent to these compounds and the potential correlation with the onset of a large anomalous Hall response. To understand the impact of the CDW order on the electronic structure in these systems, we present quantum oscillation measurements on single crystals of CsV$_3$Sb$_5$. Our data provide direct evidence that the CDW invokes a substantial reconstruction of the Fermi surface pockets associated with the vanadium orbitals and the kagome lattice framework.  In conjunction with density functional theory modeling, we are able to identify split oscillation frequencies originating from reconstructed pockets built from vanadium orbitals and Dirac-like bands. Complementary diffraction measurements are further able to demonstrate that the CDW instability has a correlated phasing of distortions between neighboring V$_3$Sb$_5$ planes, and the average structure in the CDW state is proposed.  These results provide critical insights into the underlying CDW instability in \textit{A}V$_3$Sb$_5$ kagome metals and support minimal models of CDW order arising from within the vanadium-based kagome lattice.
\end{abstract}

\maketitle

\section{Introduction}

While kagome insulators are traditionally sought as potential hosts of quantum spin liquid states and laboratories for highly frustrated magnetism  \cite{freedman2010site,wulferding2010interplay,han2012refining,fu2015evidence,han2012fractionalized}, kagome metals are equally interesting due to their potential to host topologically nontrivial electronic states interwoven with local electronic symmetry breaking. At a single-orbital tight binding level, the kagome structural motif naturally gives rise to an electronic structure with Dirac points and a flat band that together provide the potential for an interplay between topologically nontrivial surface states and substantial electron correlation effects. A wide array of instabilities have been predicted, ranging from bond density wave order \cite{PhysRevB.87.115135,PhysRevLett.97.147202}, charge fractionalization \cite{PhysRevB.81.235115, PhysRevB.83.165118}, spin liquid states \cite{yan2011spin}, charge density waves (CDW) \cite{PhysRevB.80.113102} and superconductivity \cite{PhysRevB.87.115135,ko2009doped}.

The electron filling within the kagome framework controls the formation of a wide variety of predicted electronic instabilities.  For band fillings near $5/4$ electrons per band \cite{PhysRevB.87.115135,PhysRevB.85.144402,PhysRevLett.110.126405,barros2014exotic,feng2021chiral}, a Van Hove singularity is formed at the Fermi level due to the presence of saddle points along the zone edge. Excitations between these saddle points can lead to CDW order, and, in some limits, unconventional superconductivity.  The recently discovered class of \textit{A}V$_3$Sb$_5$ (\textit{A}: K, Rb Cs) kagome metals \cite{ortiz2019new} are potential realizations of this physical mechanism with each member exhibiting thermodynamic anomalies associated with CDW order \cite{yuxiaoKVS, zeljkovicCsV3Sb5, ortizCsV3Sb5, ortiz2020superconductivity, liang2021three, chen2021roton} followed by the onset of superconductivity at lower temperatures \cite{ortizCsV3Sb5, ortiz2020superconductivity, RbV3Sb5SC}.  While there are multiple gaps identified with both evidence of $s$-wave pairing \cite{mu2021s} and evidence of nodal quasiparticles \cite{zhao2021nodal}, the interplay between superconductivity and the CDW state can in principle lead to unconventional behavior even in a fully gapped superconducting state \cite{gu2021gapless}.

The CDW instability in \textit{A}V$_3$Sb$_5$ compounds seemingly competes with superconductivity \cite{du2021pressure, chen2021double} and presages the formation of a potentially unconventional superconducting ground state \cite{ortiz2020superconductivity,chen2021double}. However, the microscopic origin of the CDW remains an open question.  Concomitant to the onset of CDW order, an exceptionally large anomalous Hall effect (AHE) appears \cite{YangKV3Sb5science, CsV3SB5hall}, despite the absence of detectable local moments or magnetic correlations \cite{2020grafKV3Sb5}.  While the normal state electronic structure is a $\mathbb{Z}_2$ topological metal \cite{ortizCsV3Sb5,ortiz2020superconductivity} and topologically-protected surface states are predicted close to the Fermi level \cite{ortizCsV3Sb5}, below the CDW transition recent scanning tunneling microscopy (STM) data \cite{yuxiaoKVS} and theoretical proposals \cite{feng2021chiral} have suggested the formation of a chiral CDW order parameter. This chiral CDW, endemic to the kagome lattice, is proposed to break time reversal symmetry and generate a large Berry curvature, potentially accounting for the AHE.  To date, however, data directly linking the onset of CDW order with reconstruction of vanadium orbitals associated with the kagome lattice in \textit{A}V$_3$Sb$_5$ is lacking. Similarly, the applicability of minimal, single orbital tight binding kagome models in multiband \textit{A}V$_3$Sb$_5$ compounds remains an open question.

\begin{figure}
\includegraphics[width=\columnwidth]{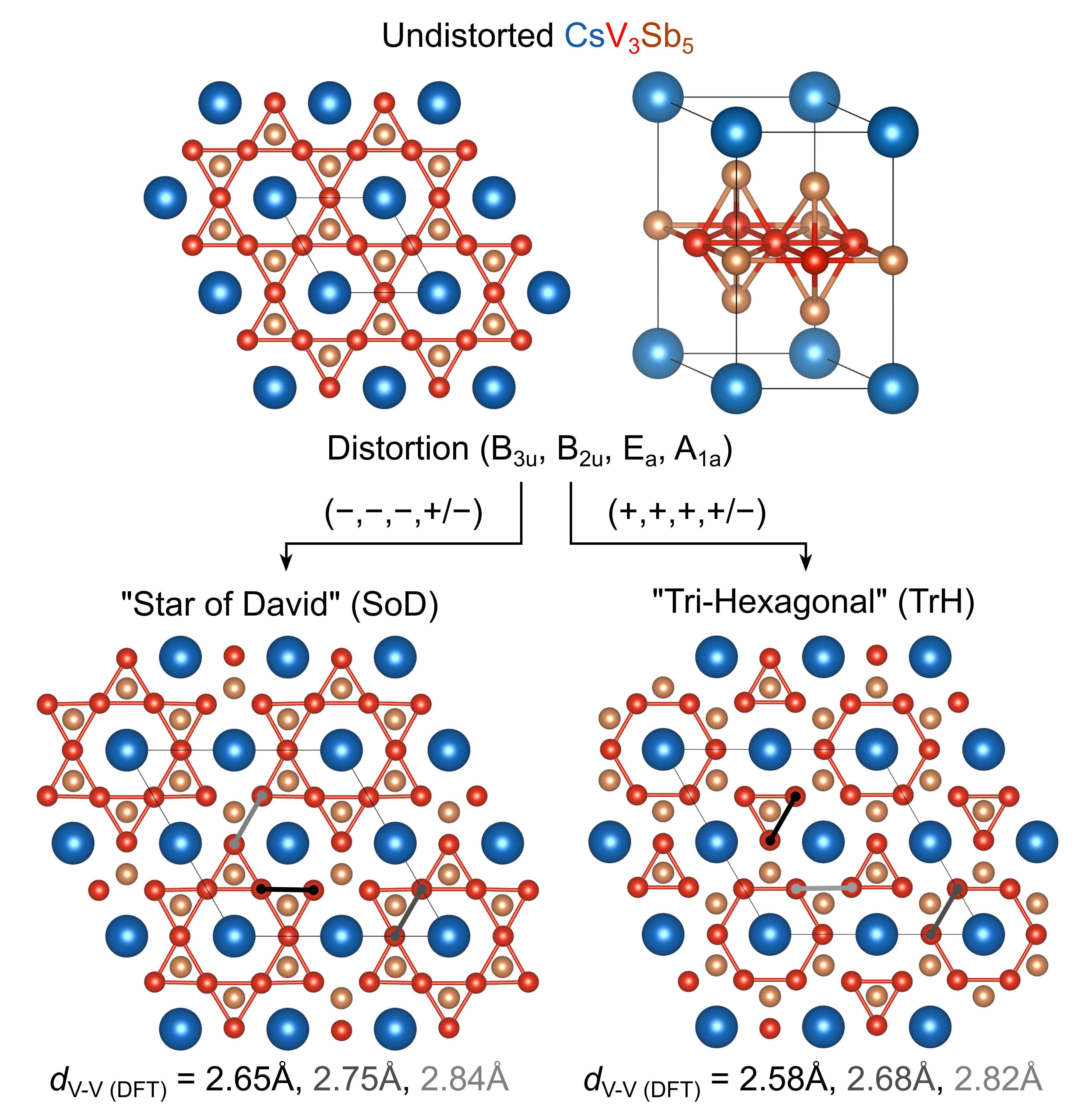}
\caption{CsV$_3$Sb$_5$ is a layered, exfoliatable, kagome metal consisting of a structurally perfect lattice of vanadium at room temperature. Upon cooling below $T^*=93$\,K, CsV$_3$Sb$_5$ exhibits charge density wave order. A concurrent structural distortion emerges as well, which is suspected to be related to the kagome ``breathing mode.'' Upon distortion and relaxation in both the positive and negative displacements, the ``breathing mode'' gives rise to the ``Star of David (SoD)'' and ``Tri-Hexagonal (TrH)'' candidate structures. }
\label{fig:structure}
\end{figure}

Specifically, STM and diffraction experiments have observed charge order with an in-plane \textbf{q}~$=(0.5, 0.5)$ wave vector in KV$_3$Sb$_5$ \cite{yuxiaoKVS} and CsV$_3$Sb$_5$ corresponding to  3\textbf{Q} CDW order. A kagome ``breathing" mode can give rise to candidate distortions such as the ``Star of David" (see Figure \ref{fig:structure}) and its inverse structure \cite{CDWKagomeMetals}, and recent studies have shown strong electron-phonon coupling in KV$_3$Sb$_5$ promoting such a distortion \cite{OpticalDetectionCDW}.  Native electronic instabilities promoting CDW order along this wave vector have long been predicted in Kagome models at select fillings near Van Hove singularities \cite{PhysRevB.87.115135,PhysRevB.85.144402,PhysRevLett.110.126405,barros2014exotic,feng2021chiral}, suggesting that a minimal model built around the kagome planes of these materials may capture the essential physics governing their unconventional electronic properties.  Notably, additional features such as unidirectional charge stripe order \cite{zeljkovicCsV3Sb5} also seemingly coexist with the 3\textbf{Q} CDW state, further connecting the underlying interactions to stripe/nematic instabilities predicted within a kagome network \cite{feng2021chiral}. 

Here we investigate the origins of the 3\textbf{Q} CDW order in \textit{A}V$_3$Sb$_5$ kagome compounds via study of Shubnikov-De Haas (SdH) quantum oscillations in magnetotransport data of CsV$_3$Sb$_5$, which has the most pronounced CDW instability within the family. By correlating quantum oscillation data with DFT models of the breathing distortion of the kagome lattice, we are able to observe the effect of the CDW on the electronic structure. Specifically, we demonstrate that a series of low-frequency quantum oscillations originate from CDW-reconstructed vanadium orbitals and exhibit transport consistent with the Dirac-like features (high mobility, low cyclotron mass) of the kagome lattice. The multiplicity and frequencies associated with these vanadium orbits are shown to originate from a reconstructed Fermi surface with small pockets linked to folded, vanadium-dominated bands.  We further demonstrate that the CDW instability is three-dimensional in nature with a resulting 2$\times$2$\times$4 superstructure.  Synchrotron x-ray data are ananlyzed to provide a model for the average superstructure.  Together, our results provide direct evidence that the in-plane CDW is derived from vanadium orbitals which comprise the kagome lattice in \textit{A}V$_3$Sb$_5$ and validate recent efforts to map the core interactions in these materials to minimal tight-binding models built from a two-dimensional kagome network. 

\section{Methods}
\subsection{Synthesis}

Single crystals of CsV$_3$Sb$_5$ were synthesized from Cs (liquid, Alfa 99.98\%), V (powder, Sigma 99.9\%) and Sb (shot, Alfa 99.999\%). As-received vanadium powder was purified in-house to remove residual oxides. Due to extreme reactivity of elemental Cs, all further preparation of CsV$_3$Sb$_5$ was performed in an argon glovebox with oxygen and moisture levels $<$0.5\,ppm. Single crystals of CsV$_3$Sb$_5$ were synthesized using the self-flux method. The flux is a eutectic mixture of CsSb and Cs$_3$Sb$_7$\cite{sangster1997cs} mixed with VSb$_2$. Elemental reagents were milled in a pre-seasoned tungsten carbide vial to form a composition which is 50\,at.\% Cs$_{0.4}$Sb$_{0.6}$ eutectic and approximately 50\,at.\% VSb$_2$. Excess antimony can be added to the flux to improve volatility if needed. The fluxes were loaded into alumina crucibles and sealed within stainless steel jackets. The samples were heated to 1000\degree C at 250\degree C/hr and soaked there for 24\,h. The samples were subsequently cooled to 900\degree C at 100\degree C/hr and then further to 500\degree C at 2\degree C/hr. Once cooled, the crystals are recovered mechanically. Crystals are hexagonal flakes with brilliant metallic luster. Samples can range up to 1\,cm in side length and up to 1\,mm thick. Elemental composition of the crystals was assessed using energy dispersive x-ray spectroscopy (EDX) using a APREO C scanning electron microscope.

\subsection{Electrical transport measurements}

Electronic transport measurements were performed using a Quantum Design 14~T Dynacool Physical Property Measurement System (PPMS). A Quantum Design rotator option was used to collect angle-dependent and temperature-dependent data. Crystals are exfoliated to remove any surface contaminants, and electrical contacts were made in a standard 4-point geometry using gold wire and silver paint. Crystals were initially mounted such that the c-axis was parallel to the field (flat plates mounted flush on resistivity stage). An alternating current of 8\,mA and 12.2\,Hz was driven in the $ab$-plane. 


\subsection{Electronic structure calculations}
DFT simulations of the electronic structure of CsV$_3$Sb$_5$ unit cell were performed in VASP \textit{v}5.4.4 using identical parameters to several recently reported studies \cite{ortizCsV3Sb5,ortiz2020superconductivity,2020MazProximitized}. We employed the PBE functional \cite{Perdew1996} with D3 corrections \cite{Grimme2011}, a 500\,eV plane wave energy cutoff, a $\Gamma$-centered $11\times11\times5$ $k$-mesh, and the recommended PAW pseudopotentials for $v$5.2. Spin orbit coupling was activated for all calculation steps except for structural relaxation. All calculations were completed with an energy convergence cutoff of 10$^{-6}$\,eV or better. The unit cell was relaxed, as previously described \cite{ortizCsV3Sb5}, with final $a$ and $c$ lattice parameters of 5.45\,\AA\ and 9.35\,\AA; in good agreement with the room-temperature values determined by X-ray diffraction, 5.52\,\AA\ and 9.36\,\AA, respectively. 

\textsc{Wannier90} \cite{Mostofi2014} was used to fit Wannier functions (Cs s, p; V s, p, d; Sb s, p; with a frozen fitting window $E_\text{F}\pm2$\,eV) and interpolate unit cell Fermi surfaces on a $101\times101\times101$ grid. Extremal orbits were determined using \textsc{Serendipity},\footnote{\textsc{Serendipity} is in development, pending an open-source release. Interested parties can contact Samuel Teicher: steicher@ucsb.edu} a new code that builds on the algorithms developed by Rourke and Julian\cite{rourke2012numerical} with additional symmetry and interactive visualization tools enabled by the python packages \textsc{spglib},\cite{togo2018spglib} \textsc{trimesh},\cite{trimesh} and \textsc{plotly}\cite{plotly}.

\begin{figure*}[t]
\centering
\includegraphics[width=7.05in]{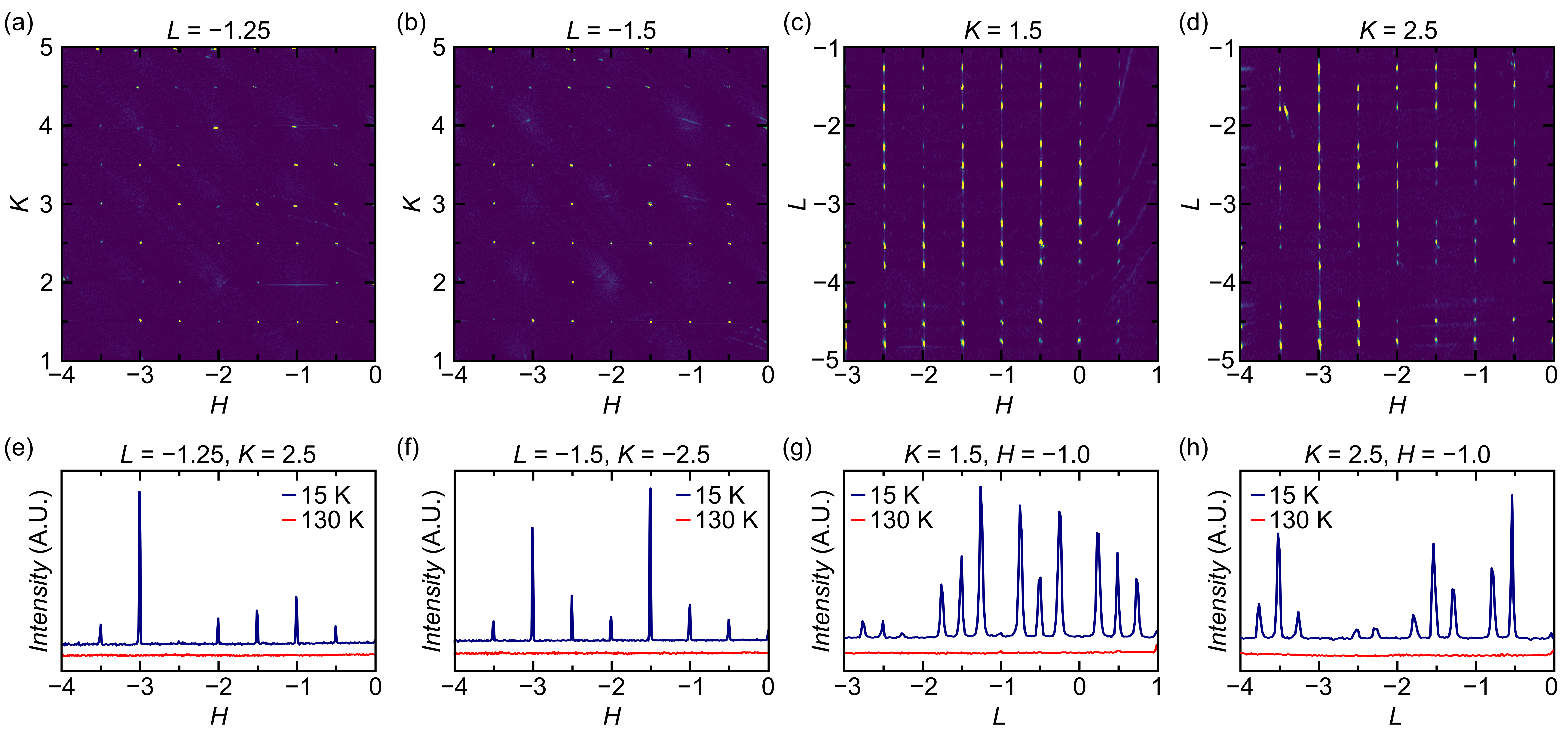}
\caption{(a--d) 2-dimensional slices through reciprocal space on half-integer Bragg planes at 15\,K, highlighting superlattice peaks in CsV$_3$Sb$_5$. (e--h) Line cuts through the 2D data, highlighting the periodicity of the superlattice peaks. Blue data are 1D traces showing data collected at base temperature (15\,K). Red data shows the data collected at 130\,K data, confirming that the superlattice peaks emerge alongside the charge density wave order emerges $T^* \sim 94$\,K. Our data indicates that the superlattice in CsV$_3$Sb$_5$ is described by a wave vector of (0.5, 0.5, 0.25).}
\label{fig:Scattering}
\end{figure*}

Supercell calculations were completed on a $2\times2\times1$ supercell simulated using identical parameters and a $5\times5\times5$ $k$-mesh that was distorted along the M$_1^+$ P$_3$ irreducible representation in \textsc{ISODISTORT}.\cite{isodistort_website,campbell2006ISODISPLACE}  M$_1^+$ P$_3$ involves four distinct distortion modes, including B$_{3u}$ and B$_{2u}$ V sublattice modes and E$_a$ and A$_{1a}$ Sb$_2$ (antimonene layer) sublattice modes. Cs and Sb$_1$ (kagome layer) atomic positions are unaffected. Negative magnitudes of the B$_{3u}$, B$_{2u}$, and E$_a$ modes recreate the ``Star of David'' (SoD) while positive magnitudes recreate the inverse ``Tri-Hexagonal" (TrH) structure. We initialized super cell structural relaxations with four different sets of mode magnitudes, (B$_{3u}$, B$_{2u}$, E$_a$, A$_{1a}$)=0.3\,\AA$\cdot$\{($+$,$+$,$+$,$+$); ($+$,$+$,$+$,$-$); ($-$,$-$,$-$,$+$); ($-$,$-$,$-$,$-$)\}, testing both the SoD and TrH structures, while additionally trialing both positive and negative magnitudes of the A$_{1a}$ mode, which corresponds to $c$-axial buckling of the antimonene lattice. 

Relaxation proceeded in three steps: volumetric optimization, followed by adjustment of the ionic positions, and finally a free relaxation of the super cell lattice parameters and ionic positions simultaneously. Ultimately, the two SoD configurations relaxed to a similar structure, and the same was true for the two TrH structures. In the final SoD structure we find a positive A$_{1a}$ mode, corresponding to Sb$_2$ atoms moving further from the smaller triangles and closer to the larger triangles of the SoD structure. In the final TrH structure, we  find that a negative A$_{1a}$ mode is favored, in which the Sb$_2$ atoms similarly move away from the smaller V-V triangle units. In each case, $c$-axial buckling of the Sb$_2$ layer is minuscule, 0.005$c$ and 0.002$c$ for SoD and TrH, respectively. 

The SoD and TrH distorted super cells are energetically favored over the undistorted unit cell by 4.7\,meV and 13.5\,meV per formula unit, respectively, consistent with recently reported simulations \cite{CDWKagomeMetals}. Super cell band unfolding employed a modified version of \textsc{VaspBandUnfolding} \cite{vaspbandunfolding}. Unfolded Fermi surface slices in Fig. \ref{fig:DistortedFS} were calculated on a $51\times51$ BZ mesh. Cubic spline interpolation was used for smoothing/upsampling prior to projecting onto the larger display range.

Fermi levels for the electronic structure calculations in Fig. \ref{fig:Bands} and Fig. \ref{fig:DistortedFS} were determined based on prior experiment. Additional discussion is provided in the supporting material \cite{ESI}. Errors in extracting frequencies associated with extremal orbits of the unfolded supercells were determined by graphically selecting orbit paths clearly within a given orbit and those clearly outside and are a product of the pixel resolution of the calculations. The top of the error bar is the outer bounding area, the bottom the inner bounding area, and the average of these bounds was chosen as the nominal value. Errors in determining the extremal orbits of the parent structure are small and not shown ($<$10\,T).

\subsection{X-ray diffraction measurements}
High dynamic range x-ray diffraction maps were collected at the QM2 beamline at CHESS. The incident x-ray wavelength was 0.42755\AA, selected using a double-bounce diamond monochromator. Temperature was controlled by bathing the small single crystal samples inside a stream of cold flowing helium gas. Diffraction was recorded in transmission though the sample using a 6 megapixel photon-counting pixel-array detector with a silicon sensor layer. Full 360 degree sample rotations, sliced into 0.1 degree frames, were indexed to the high-temperature crystal structure and transformed to reciprocal space. Some elements of the data reduction employed the NeXpy software package.  Crystal structures were visualized in \textsc{VESTA}.\cite{Momma2011}.  Diffraction data were analyzed within the APEX3 software package and data were corrected for absorption and extinction effects. Refinement of the structure was performed using the integrated SHELX software package.\cite{sheldrick2015crystal}  Charge flipping simulations of diffraction data were performed using the TOPAZ software package.\cite{oszlanyi2004ab,oszlanyi2005ab,coelho2007charge}

\subsection{Second harmonic generation optical measurements}
Second harmonic generation (SHG) measurements were performed using an ultrafast laser with a pulse duration of 40 fs and a repetition rate of 50 kHz. The laser was tuned to a center wavelength of 800 nm and a sample fluence of 3 mJ/cm$^2$. An oblique incidence reflection geometry was employed with both incoming and outgoing beams P-polarized. The reflected SHG at 400 nm was isolated with a spectral filter and detected using a back-illuminated CMOS image sensor. Overall SHG intensities were extracted by averaging over the scattering plane angle. A sample-in-vacuum optical cryostat was used to cool the sample below the CDW phase transition temperature.

\section{Experimental Results}

\subsection{Crystal Structure}

The \textit{A}V$_3$Sb$_5$ (\textit{A}: K, Rb Cs) family of kagome metals are layered, exfoliable materials consisting of V$_3$Sb$_5$ slabs intercalated by alkali metal cations. The vanadium sublattice forms a perfect kagome lattice under ambient conditions (Figure \ref{fig:structure}). CsV$_3$Sb$_5$ is the terminal endpoint of the alkali-metal series, shows the highest superconducting transition ($T_c = 2.5$\,K), and an onset of CDW order below $T^* = 94$\,K \cite{ortiz2020superconductivity}. The CDW is accompanied by a weak structural distortion manifest as a superlattice of Bragg scattering in synchrotron x-ray diffraction data.

Early measurements within the $L=0$ scattering plane resolved only \textbf{q}~$=(0.5, 0, 0)$ and $(0, 0.5,0)$-type superlattice reflections \cite{ortizCsV3Sb5}, which is seemingly at odds with recent STM reports of 2$\times$2 supercells associated with 3$\bf{Q}$ charge order.  To address this, an expanded exploration of superlattice peaks was conducted at finite $L$-values with the results shown in Fig. \ref{fig:Scattering}.  In this higher resolution data,  a more complex, three-dimensional superlattice structure is observed that is best indexed by a \textbf{q}~$=(0.5, 0.5, 0.25)$ wave vector.  (0.5, 0.5)-type superlattice reflections are largely not resolvable in the $L=0$ plane, accounting for the initial failure to index them.  The in-plane component of the superlattice modulation agrees with the 3\textbf{Q} structure observed in local probes. The superlattice peaks at (0.5, 0.5, 0.25)-type positions vanish above the CDW ordering temperature and indicate a modulation of the in-plane distortion along the $c$-axis (interplane phasing).  

Considering first the in-plane distortions allowed on an idealized kagome lattice, the kagome ``breathing'' mode often leads to lower energy structures, and this mode matches preliminary conclusions drawn from STM and DFT studies of KV$_3$Sb$_5$ \cite{yuxiaoKVS,CDWKagomeMetals,OpticalDetectionCDW}. As shown in Figure \ref{fig:structure}, the structure can distort between two potential candidates: (1) the SoD distortion and (2) the TrH distortion. The phasing along the c-axis, which governs the modulation of the distortion motifs along the out-of-plane direction are naively expected to be of a lower energy scale than the in-plane components.  To determine the nature of the three-dimensional superstructure that forms below the CDW transition, the low-temperature (15 K) diffraction data was refined.  Approximately 30000 reflections ($\sim$4500 unique) were indexed within a hexagonal unit cell with lattice parameters $a=b=11.05410(13)$\AA, $c=37.334(5)$\AA, and $\alpha=\beta=90^{\circ}$ $\gamma=120^{\circ}$.

\begin{figure}
	\includegraphics[width=\columnwidth]{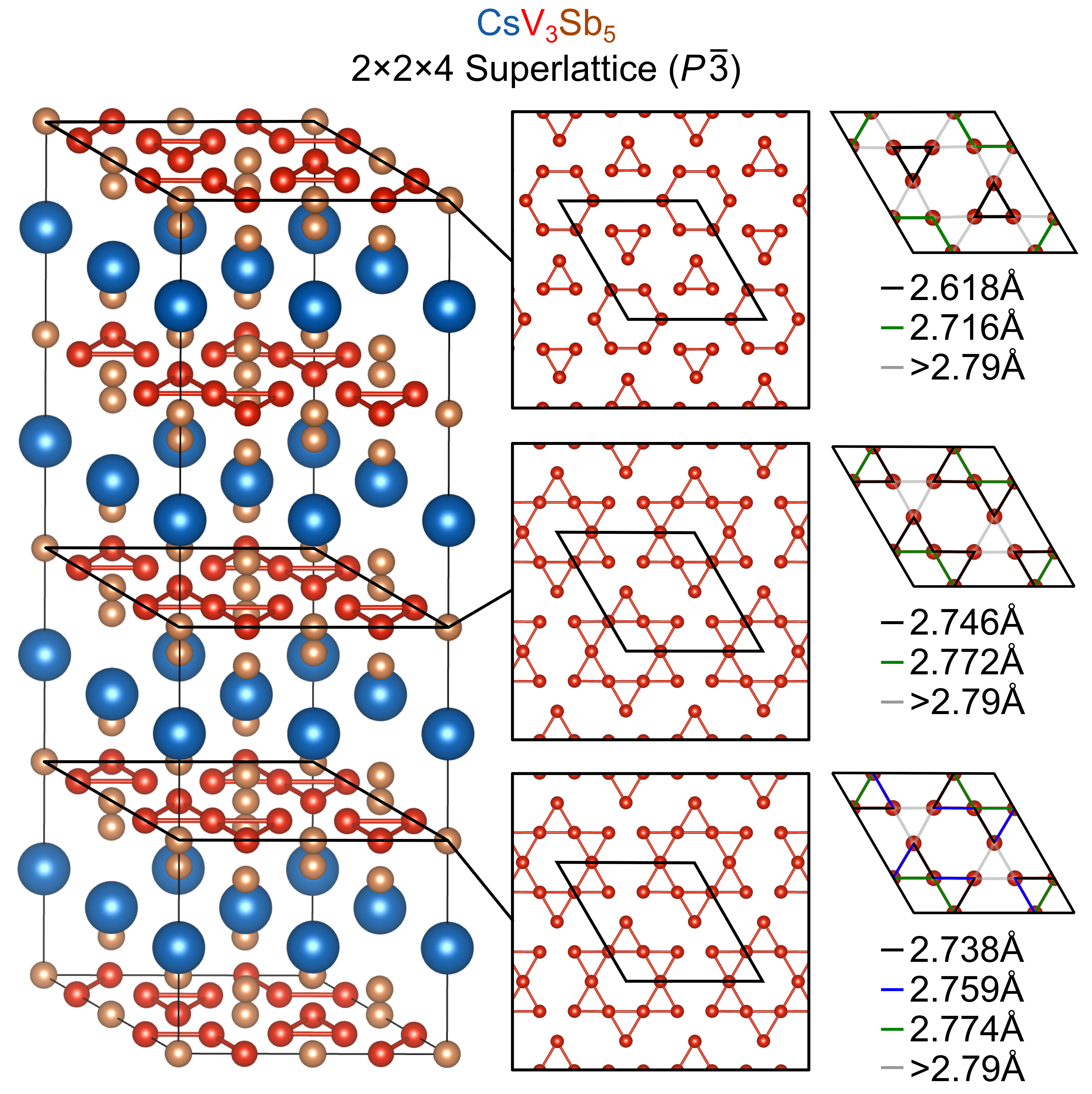}
	\caption{Single-crystal diffraction data implies that CsV$_3$Sb$_5$ distorts into 2$\times$2$\times$4 supercell, where the kagome layers exhibit both TrH- and SoD-like distortions. Our current model indexes the cell in the $P\bar{3}$ space group. The SoD-like distortions are substantially weaker than the TrH-like distortion, and two unique SoD layers are noted.}
	\label{fig:CDWstructure}
\end{figure}

SHG data indicate that inversion symmetry is not broken below the ordering transition \cite{ESI}, and, as a result, data were initially analyzed via charge-flipping in the $P\bar{1}$ space group.  This provided a baseline visualization for distortions below the transition, and, absent any further constraints, already suggests an average cell with modulation between TrH and SoD-type distortions along the c-axis.  Further refinement was then pursued within the space group $P\bar{3}$, assuming a minimal three-fold symmetry that conformed with the diffraction data as well as inversion symmetry demonstrated from the SHG data. While the data can potentially be indexed in a lower rotational symmetry (e.g. centered orthorhombic), we could not find sufficient evidence within the current data to perform the refinement lower than $P\bar{3}$. For the solution in $P\bar{3}$, a twinning model was used with twins realized via a two-fold rotation along the (0, 0, 1) axis. We qualitatively tested alternate hexagonal twinning models, and in all cases the refined structure remains nearly unchanged. 

Atomic positions and displacement parameters were refined in $P\bar{3}$ (R1 = 0.089, GoF = 1.32) with the resulting refinement parameters provided in the supplemental information \cite{ESI}.  The resulting $P\bar{3}$ structure is shown in Fig. \ref{fig:CDWstructure}. To highlight the differences in bonding and V-V motifs, we have selected to draw V-V bond lengths $\leq2.79$\AA. The middle panels demonstrate the different motifs with distortions in each kagome plane highlighted.  The top and bottom layers of the lattice assume an in-plane TrH distortion while the intervening layers assume a weak SoD-like distortion. The right most panels of Figure \ref{fig:CDWstructure} identify the distorted V-V bond distances. For graphical simplicity, bonds within 0.0025\AA\, of their mean value were grouped and averaged. Full bonding information is available in the CIF file.\cite{ESI} The TrH-like layers feature the most distinct distortion, which manifests the largest deviation in bond lengths from the parent structure. The two unique SoD-like layers are similar, though the central layer exhibits slightly weaker V-V bond distortions.  We emphasize here that this is a depiction of the \textit{average} structure produced by modeling the x-ray diffraction data.  More complex twinning effects or stacking disorder within the 4 layer unit cell can influence the appearance of the average structure.  

\subsection{Electronic Structure}

As the interlayer interactions are expected to be weak in CsV$_3$Sb$_5$, we neglect the impact of the $c$-axis component of the superlattice on the electronic structure and focus on the impact of the in-plane distortion modes. This was verified by comparing the calculated band structures of the nominal $2\times 2\times 1$ cell with the $2 \times 2 \times 4$ cell proposed by SCXRD \cite{ESI}.  Candidate structures (M$_{1}^+$ P$_3$ irrep.) matching the pure TrH and SoD distortions were selected for DFT relaxation and band structure calculations. Our DFT studies find that both the SoD and TrH distortions are slightly favored over the undistorted structure; by 4.7\,meV/f.u and 13.5\,meV/f.u, respectively. DFT-relaxed structures are shown alongside the experimental, undistorted crystal structure in Figure \ref{fig:structure}. 


Despite the low stabilization energy of the distorted structures relative to the parent structure, the predicted vanadium lattice distortions are significant. The V-V bond lengths are all of equal length (2.72\AA)\, in the parent structure, and transform to 3 distinct lengths: 2.65\AA\, 2.75\AA\, and 2.84\AA\ for the SoD distortion, and 2.58\AA\, 2.68\AA\, and 2.82\AA\, for the TrH distortion. The experimentally refined structure shows slightly weaker distortions and corresponding lengths of 2.74\AA\, 2.77\AA\, and 2.79\AA\ for SoD layers and 2.62\AA\, 2.72\AA\, and 2.86\AA\ for the TrH layers. Concurrently, the Sb graphitic sublattice is fragmented into individual hexagons and also hosts a slight buckling in the $c$-direction for both distorted structure types.


\begin{figure}
\includegraphics[width=\columnwidth]{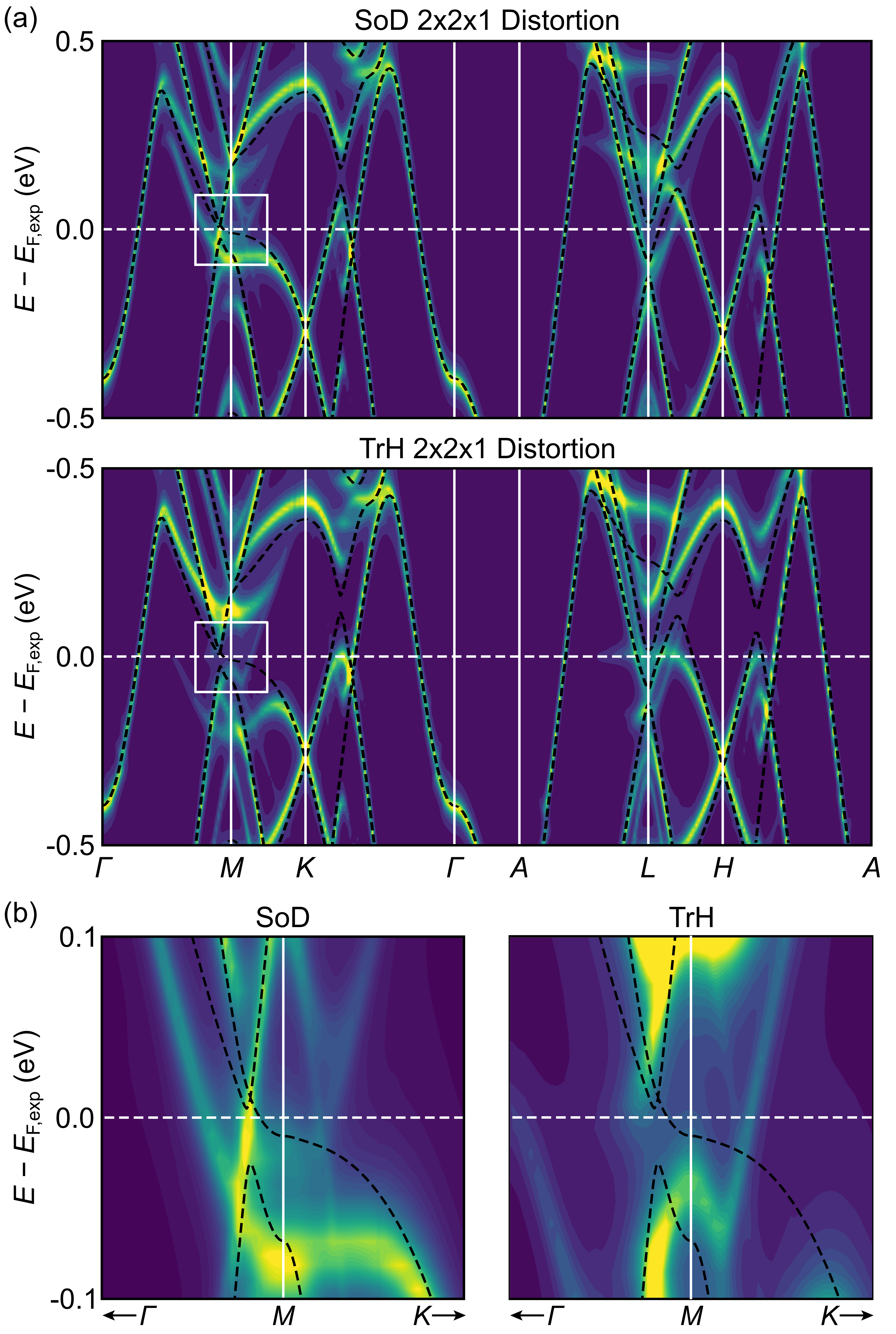}
\caption{(a) Unfolded electronic structure of the ``Star of David (SoD)'' and ``Tri-Hexagonal (TrH)" distortions in CsV$_3$Sb$_5$ with the undistorted electronic structure superimposed for comparison (black). Largest perturbation to the structure appears near the Dirac-like bands near M. (b) Close-up of the changes near the M-point, highlighting the new bands that appear as a result of the CDW and the associated structural distortion.}
\label{fig:Bands}
\end{figure}

Figure \ref{fig:Bands} shows the effect of the two superlattice types (SoD and TrH) on the \textit{ab initio} electronic structure of CsV$_3$Sb$_5$. The resulting band diagrams were unfolded for comparison to the undistorted electronic structure shown in previous works \cite{ortiz2019new,ortizCsV3Sb5}. The heat map shows the relative projections of the electronic states after the unfolding, and the black bands are the undistorted structure of CsV$_3$Sb$_5$. In the low-temperature distorted state, the electronic structure is largely unperturbed, particularly the central band about $\Gamma$ which derives from the Sb $p$-orbitals. However, the bands near the M-point, which are the relevant Dirac-like bands associated with the vanadium $d$-orbitals, are altered significantly. 

Figure \ref{fig:Bands}(b) shows expanded views of the SoD and TrH electronic structures in the vicinity of the $M$-points. An orbital decomposed band diagram (``orbital bands'') further identifies these states as originating primarily from vanadium orbitals \cite{ESI}.  The emergence of the CDW and the resulting superlattice therefore has a clear effect on the electronic structure near $E_\text{F}$ for the vanadium orbitals comprising the Dirac-like crossings. This effect has a significant impact on the Fermi surface and is expected to impact transport sensitive to topographical changes at $E_\text{F}$. Experimental detection of these effects are discussed in the next section.

\begin{figure*}[t]
\centering
\includegraphics[width=7.05in]{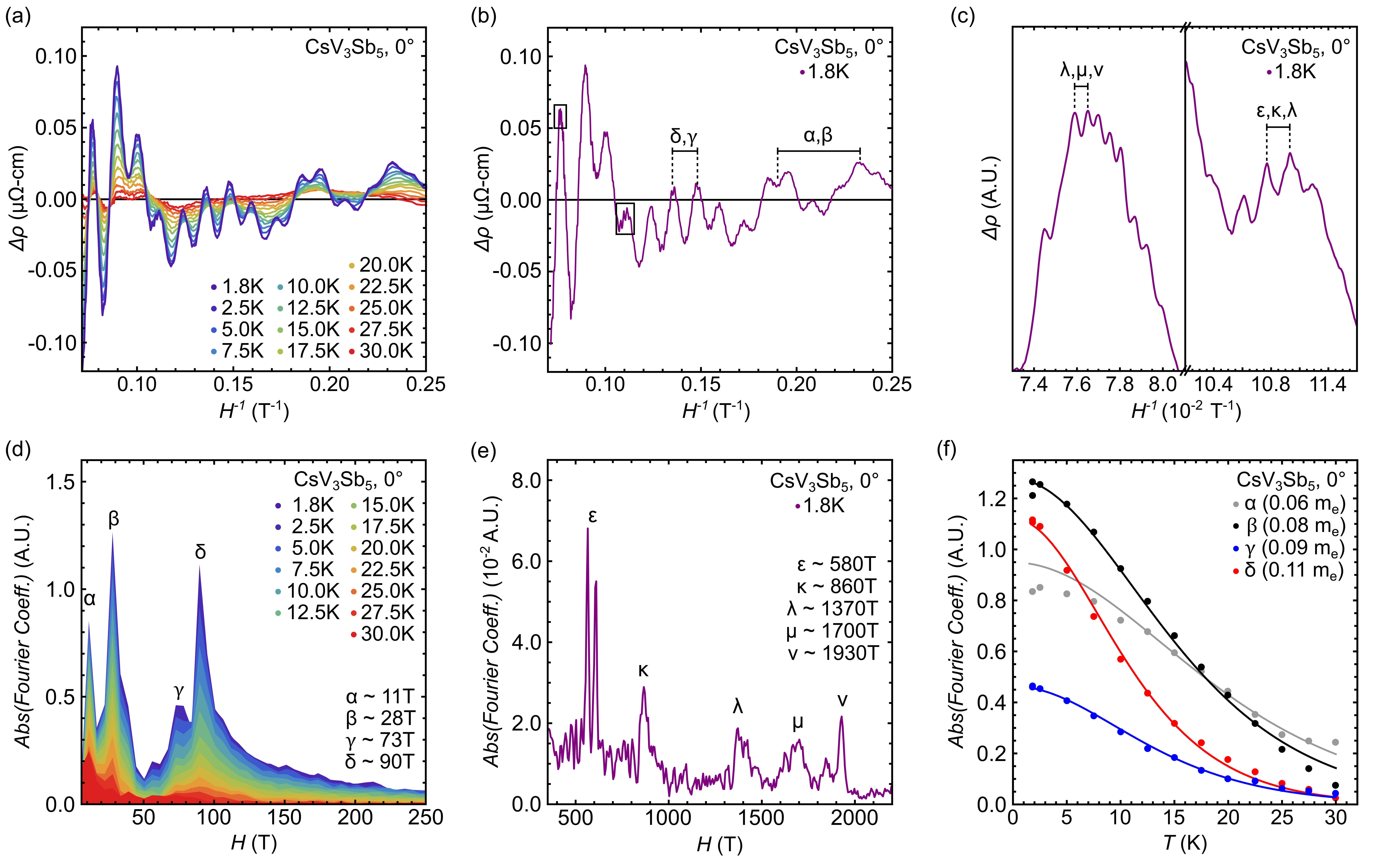}
\caption{(a) Oscillatory component of the magnetoresistance extracted from the temperature-dependent SdH data collected on single crystals of CsV$_3$Sb$_5$ mounted with the c-axis parallel to the magnetic field (0\degree). (b) A high-resolution scan at 1.8\,K shows significant contributions from high-frequency modes, particularly at the peaks and troughs of the general oscillatory behavior. (c) Magnified view of the high-frequency features, providing visual confirmation for multiple high-frequency modes. (d,e) Fourier transformation of the quantum oscillation data, showing the low- and high-frequency components of the power spectrum. The 9 unique frequencies have been assigned greek letters. (f) Magnitudes of the Fourier coefficients, which are used in conjunction with the Lifshitz-Kosevich (LK) formula to extract the cyclotron ``effective masses.'' All low-frequency modes show very low effective masses, consistent with transport originating from the Dirac-like crossings at M.}
\label{fig:SdHTemp}
\end{figure*}

\subsection{Quantum Oscillation Measurements}
A effective bulk probe of the low energy band structure is the measurement of quantum oscillations in high-field electron transport measurements.  Crystals of the \textit{A}V$_3$Sb$_5$ kagome metals are high mobility metals with low residual resistivity ($\sim0.1~\mu\Omega$-cm) values \cite{ortiz2019new,ortiz2020superconductivity,YangKV3Sb5science}, rendering quantum oscillation measurements an appealing probe for exploring the low temperature electronic structure. 

\begin{figure*}[t]
\centering
\includegraphics[width=7.05in]{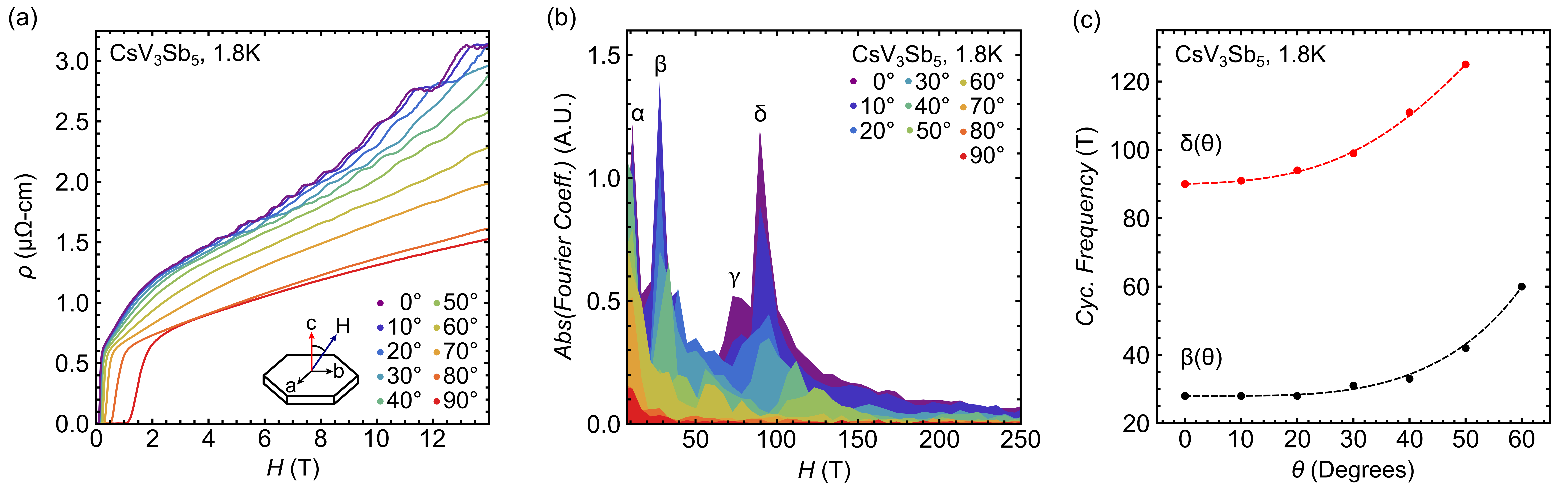}
\caption{(a) Raw electronic resistivity as a function as field and angle, where the angle is defined as between the c-axis and the magnetic field. Oscillations are clearly visible $>$2\,T. The oscillations appear dampened with angle, vanishing for $\theta>$60\degree. (b) Fourier transform of angle-resolved SdH data, showing primary low-frequency contributions to the power spectrum. The high-frequency contributions are suppressed rapidly by rotation. (c) The most prominent frequencies ($\beta$, $\delta$) are shown as a function of rotation angle -- dashed lines serve as a guide to the eye. The frequencies were estimated from (b) using Gaussian functions to approximate both the broadening and shortening of peaks. These frequencies exhibit a delayed onset of the angular dependence, consistent with Dirac-like pockets.}
\label{fig:SdHAngle}
\end{figure*}

Figure \ref{fig:SdHTemp} presents a series of temperature-dependent quantum oscillation measurements on stoichiometric crystals of CsV$_3$Sb$_5$ with RRR~$\approx~80$. The crystals were mounted with the $c$-axis parallel to the magnetic field while current was driven within the $ab$-plane. The normal ``background'' magnetoresistance (MR) was modeled using a power function $\rho_0 \approx \alpha H^\beta + \gamma$ fit over the range from 4\,T to 14\,T. The oscillatory component of the MR was then isolated by subtracting the background MR $\Delta \rho(H) = \rho(H) - \rho_0(H)$. Figure \ref{fig:SdHTemp}(a) shows the oscillatory component of the magnetoresistance as a function of temperature and field, and oscillations are seen to persist up to $25$\,K. The oscillation pattern is relatively complex, with multiple harmonics visible by inspection. 

Quantum oscillation data collected at 1.8\,K is isolated in Figure \ref{fig:SdHTemp}(b), where multiple frequency components have been noted by Greek letters. At higher fields, we further highlight several regions that show contributions from additional, higher frequency oscillations in Figure \ref{fig:SdHTemp}(c). Due to the presence of multiple closely spaced frequencies (discussed in the next paragraph), we have grouped similar frequency components together in this initial inspection of the subtracted data. All frequencies persist between different measurements and different crystals. 

Turning first to the low frequency spectrum, Figure \ref{fig:SdHTemp}(d) shows the Fourier transform of the data at multiple temperatures with $H\leq250$\,T. Four well-defined frequencies appear ($\alpha, \beta, \gamma, \delta$). At higher frequencies with $400<H<2000$\,T, the Fourier transform in Figure \ref{fig:SdHTemp}(e) shows an additional five well-defined frequencies ($\epsilon, \kappa, \lambda, \mu, \eta$). The peak designated $\epsilon$ technically appears as two sharp peaks; however, this effect is likely extrinsic, and we currently consider the $\epsilon$ peak as the average of these two peaks. 

While the modes above 250\,T vanish quickly with increasing temperature above 2\,K, the low frequency modes remain well-defined up to 25\,K. The temperature dependence of the Fourier coefficients of the $\alpha$, $\beta$, $\gamma$ and $\delta$ orbits are shown in Figure \ref{fig:SdHTemp}(f). The cyclotron ``effective mass'' ($m^*_\text{eff}$) can be extracted using the approximate Lifshitz-Kosevich (LK) form $a_i(T)\approx X/(B \sinh{X/B})$, where $X=\alpha\,m^*_\text{eff}T$. Here, $B$ is the magnetic flux density and is typically selected as the mean field within the FFT window. The parameter $\alpha$ is a constant defined as 14.69\,T/K.  The resulting $m^*_\text{eff}$ values are low for these low frequency orbits---nearly $1/10$ of the free electron mass---consistent with transport originating from the Dirac modes expected near the $M$-point. 
 

Whereas the temperature-dependence of the quantum oscillations provides information regarding the scattering, lifetime, and effective mass of the carriers, the angular-dependence can provide information regarding the topography of the Fermi surface. Figure \ref{fig:SdHAngle}(a) presents a series of angle-dependent quantum oscillation measurements collected at 1.8\,K where $\theta=0^{\circ}$ denotes the $c$-axis parallel to the $H$-field. The Fourier transforms of the data in Figure \ref{fig:SdHAngle}(a) are shown in Figure \ref{fig:SdHAngle}(b), and the frequencies of the $\delta$ and $\beta$ orbits are plotted as a function of angle in Figure \ref{fig:SdHAngle}(c). The $\alpha$ and $\gamma$ orbits shift and quickly convolve into neighboring frequencies with increasing angle, precluding their analysis at finite $\theta$.

Conceptually, the orbits that generate the oscillations can be imagined as slices through the Fermi surface at different approach angles. `Extremal' cross-sections with the largest and smallest cross-sectional area will generate distinct oscillation frequencies. For example, a perfectly spherical Fermi pocket exhibits no angular dependence and only one frequency from the circular cross-section. A strongly anisotropic pocket (e.g. those from 2D Dirac cones) would show a strong dependence with angle, as oblique slices through a cylinder become progressively larger as the angle increases. The sharp upturns seen above 40$^{\circ}$ in the $\delta$ and $\gamma$ orbits are consistent with orbits derived from strongly anisotropic pockets, and---as we will demonstrate in the next section---are best ascribed to electrons within Dirac-like features associated with the vanadium kagome lattice.

\subsection{Fermi Surface Topography and Frequency Correlation}

The transport data shown in Figures  \ref{fig:SdHTemp} and \ref{fig:SdHAngle} reveal a complex superposition of quantum oscillations originating from multiple portions of the Fermi surface. In order to identify how the CDW and the associated crystallographic distortions perturb the Fermi surface, the oscillation frequencies (i.e. enclosed Fermi surface pockets) seen experimentally must be correlated to the DFT-calculated Fermi surfaces. To do so, we first examine the undistorted Fermi surface in the context of the possible extremal orbits. 

To determine the orbits accurately, the Fermi energy needs to be well-defined. It is worth taking a moment to review the spread of $E_\text{F}$'s reported in the current literature, as the Dirac-like nature of the bands near $E_\text{F}$ renders rapid changes in the sizes of electron pockets with relatively minor shifts in Fermi energy. Initial DFT studies  \cite{ortiz2019new,ortiz2020superconductivity,ortizCsV3Sb5,yuxiaoKVS,YangKV3Sb5science} found Fermi levels slightly below those determined experimentally by ARPES \cite{yan2011spin,ortizCsV3Sb5} and STM measurements \cite{yuxiaoKVS,zeljkovicCsV3Sb5}. For simplicity, we will refer to these earlier results as $E_\text{F,lit}$ and $E_\text{F,exp}$ respectively. Recent DFT studies have since provided self-consistent results closer to experimental values \cite{CDWKagomeMetals}; we refer to this value as $E_\text{F,DFT}$, which agree with the self-consistent DFT calculations presented in this work. 

\begin{figure*}[t]
\centering
\includegraphics[width=7.05in]{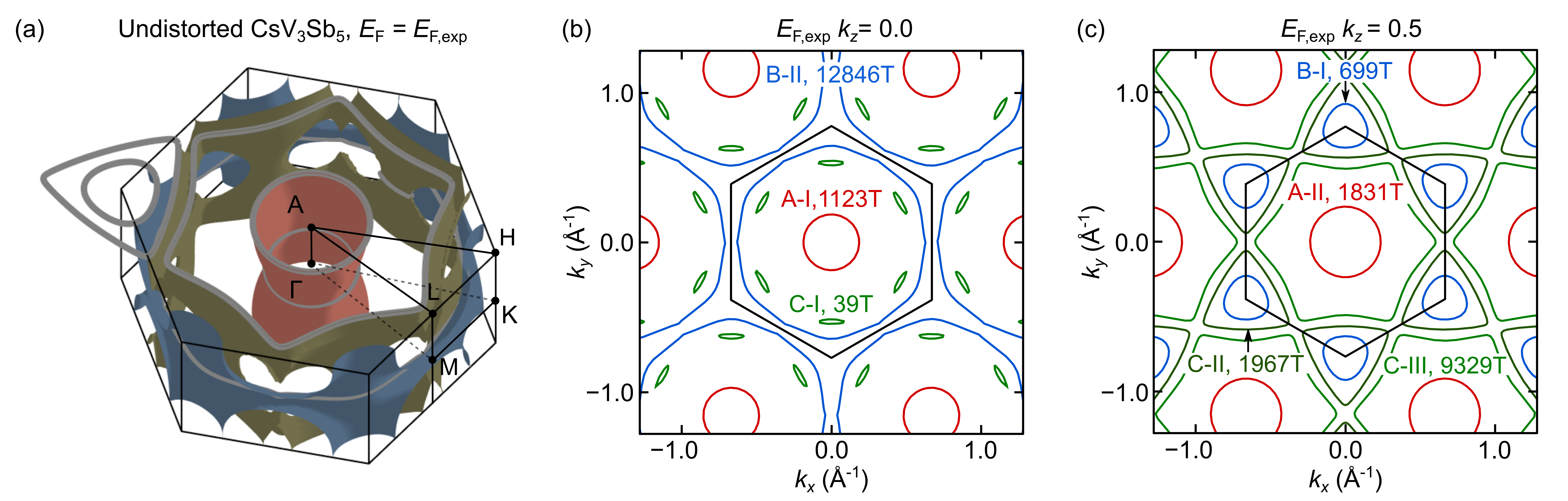}
\caption{(a) The Fermi surface for undistorted CsV$_3$Sb$_5$ calculated with $E_\text{F} = E_\text{F,exp}$ shows a variety of potential orbits. Orbits calculated by the \textsc{Serendipity} python package are shown highlighted in grey. (b,c) For clarity, isoenergy contours at $k_z = 0$ and $k_z = 0.5$ for the $E_\text{F} = E_\text{F,exp}$ surface are shown with all unique extremal orbits marked. Several high- and low-frequency orbits can be identified. The Fermi surface pocket of origin is indicated through the orbit color, which will be used for comparison throughout our discussion. While there are clearly multiple frequencies in the predicted spectrum, note that there is only 1 symmetry unique (39\,T) low frequency-mode, which seems at odds with our experimental observations. Further, for frequencies $400<f<2000$\,T, there are only 4 predicted modes, as opposed to the 5 experimentally observed components.}
\label{fig:FS}
\end{figure*}

Figure \ref{fig:FS} shows the calculated Fermi surface using the \textit{undistorted} CsV$_3$Sb$_5$ structure where $E_\text{F} = E_\text{F,exp}$. Despite already motivating that the underlying electronic structure is perturbed by the influence of the CDW, it is nevertheless instructive to first understand the undistorted Fermi surface. The orbits identified by \textsc{Serendipity} are shown in Figure \ref{fig:FS}(a) in grey. 2D slices of the Fermi surface at $k_z = 0$ and $k_z = 0.5$ are provided for a more convenient comparison, since all of the extremal orbits at $E_\text{F,exp}$ occur on these two high-symmetry planes. The Fermi surface maps reveal a variety of possible extremal orbits, confirming that the quantum oscillations in CsV$_3$Sb$_5$ should contain multiple frequencies, though the 9329\,T and 12846\,T frequencies are significantly above the experimental range of detection in \ref{fig:SdHTemp}(d,e) and Figure \ref{fig:SdHAngle}(b). 

First considering the measurable high frequencies (250\,T~$<f<$~2000\,T), we find only four extremal orbits (B-I, 699\,T; A-I, 1123\,T; A-II, 1831\,T; and C-II, 1967\,T) to match to five measured values ($\eta\approx580$\,T, $\kappa\approx860$\,T, $\lambda\approx1370$T, $\mu\approx1700$\,T, $\nu\approx1930$\,T). At lower frequencies, the agreement is much worse: only one calculated low frequency orbit (C-I, 39\,T) is found to compare with four low frequency oscillations observed in experiments ($\alpha\approx11$\,T, $\beta\approx28$\,T, $\gamma\approx74$\,T, and $\delta\approx90$\,T). A table of the calculated (DFT) frequencies at $E_\text{F} = E_\text{F,exp}$, along with the associated cyclotron masses $m_\text{cyc}$ has been included in the supplementary supporting material (SFig. 3).\cite{ESI} 


\begin{figure}[!ht]
\includegraphics[width=\columnwidth]{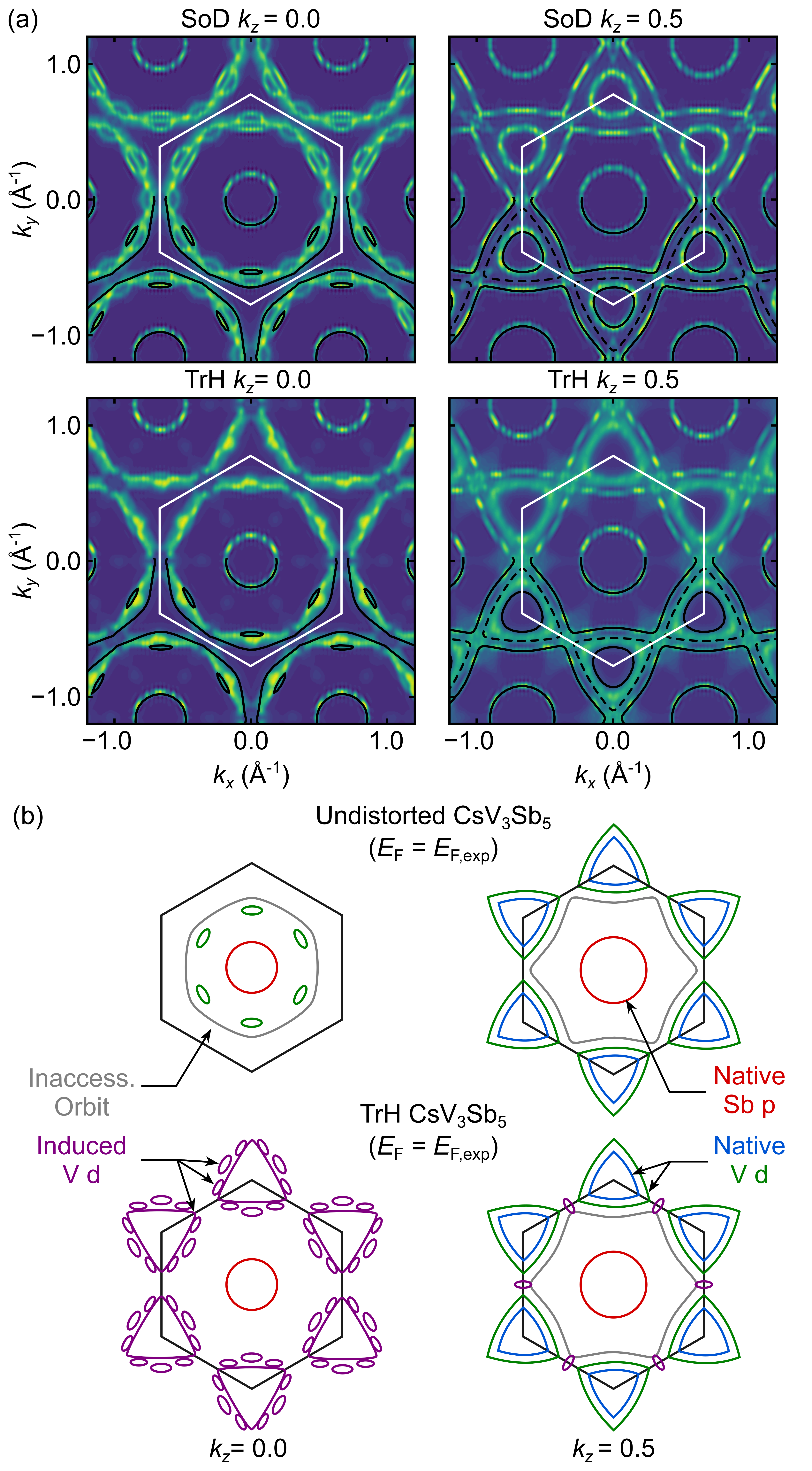}
\caption{(a) Fermi surfaces for the SoD and TrH distortions in CsV$_3$Sb$_5$ generated from the unfolded supercell band structures. Undistorted Fermi surface (black lines) contours are shown for comparison. A significant reconstruction of the V $d$ states occurs. Most high-frequency modes are preserved, though the C-II orbit (dashed) is gapped in the SoD structure. (b) Schematic description of Fermi surface reconstruction. Grey orbits are well above measurable frequencies. Band reconstruction in the TrH supercell introduces one additional triangular orbit around the $M$-point while preserving higher frequency native V $d$ and Sb $p$ orbits. Notably, the distortion also introduces 3 smaller, Dirac-like orbits (purple) shown in Figure 4 consistent with our measurements.}
\label{fig:DistortedFS}
\end{figure}

Given the significant CDW band reconstruction presented in Fig. \ref{fig:Bands}, a low-temperature Fermi surface modification in CsV$_3$Sb$_5$ is expected. Figure \ref{fig:DistortedFS} displays unfolded Fermi surface slices for pure SoD and TrH $2\times 2\times 1$ supercells at $E_\text{F,exp}$, demonstrating this reconstruction. While the central Sb $p$-orbits are largely unaffected by the CDW, V $d$-bands gap and change the Fermi surface. On the $k_z=0.0$ plane, the single, large B-II orbit reconstructs into small orbits in both structures,  generating triangular orbits around the $K$ points at the corners of the Brillouin zone. On the $k_z=0.5$ plane, the A-II central pocket and the smaller B-I orbit around $K$ are largely unaffected by the CDW; however, the larger C-II orbit (dashed line) is strongly affected. In the SoD structure, this orbit is completely gapped out at $E_\text{F,exp}$, while in the TrH structure, the C-II orbit persists. 

Supporting these models, prior ARPES results show the A-I and A-II orbits as well as the B-I orbit in KV$_3$Sb$_5$ and CsV$_3$Sb$_5$ both above and below the CDW transition \cite{ortizCsV3Sb5,yuxiaoKVS}. STM results at lower temperature also show the preservation of the A-I and A-II orbits and are consistent with at least one triangular orbit at the $K$-points \cite{zeljkovicCsV3Sb5}. Therefore, with regard to the preservation of the A-I, A-II, and B-I orbits, CDW calculations in both SoD and TrH structures appear consistent with experimental results to-date.  STM data also validate the reconstruction of the B-II orbit captured within the DFT models; however our models also predict numerous other changes in the low frequency (small orbit) regime. 

Investigating the band reconstruction in the distorted state further, the data shown in Figure \ref{fig:DistortedFS}(a) clearly show additional features. Figure \ref{fig:DistortedFS}(b) provides a simplified, pictorial representation of the Fermi surface and closed orbits in the undistorted and TrH structures to aide discussion. All possible orbits that exist in the undistorted structure \ref{fig:DistortedFS}(a) are depicted at $E_\text{F,exp}$. Orbits which are too large to be experimentally observed with our current data are shown in grey. The remaining orbits are color-coded consistent with the pocket designations shown previously in Figure \ref{fig:FS}. For this qualitative comparison, we focus on the orbits within the TrH structure for two reasons; 1) the additional modes are less obvious in the 2D data for the TrH (but no less relevant), and 2) as we will show, the TrH structure produces one additional frequency in the ``mid-frequency" regime which makes the presence of layers with this configuration distinguishable.

Focusing on comparison of the high-frequency orbits, we see that there are a total of 4 experimentally accessible orbits at $E_\text{F,exp}$ in the undistorted structure. Upon introducing the TrH distortion, several key changes occur. In the $k_z = 0$ plane, the distortion generates 3 distinct vanadium $d$-orbits at $E_\text{F,exp}$ by shifting and gapping bands around the $M$ and $K$ points near the corners and sides of the zone. Notably, band reconstruction about the $M$ point forms a new medium-frequency triangular orbit at $E_\text{F,exp}$.  The $k_z = 0.5$ plane, in contrast, is largely preserved in the new configuration, with the exception of the addition of one Dirac-like orbit at the $L$ point. 


Thus, in this high frequency regime, the TrH distortion has several effects: 1) the generation of 3 additional Dirac-like modes, 2) the preservation of high frequency orbits primarily comprised of Sb-states, and 3) the introduction of a new triangular orbit from the band reconstruction. It is worth noting that other, smaller Dirac-like orbits are likely present, but the resolution of the present supercell calculation limits our search to orbits $>$30\,T.

\begin{figure}[t]
\includegraphics[width=\columnwidth]{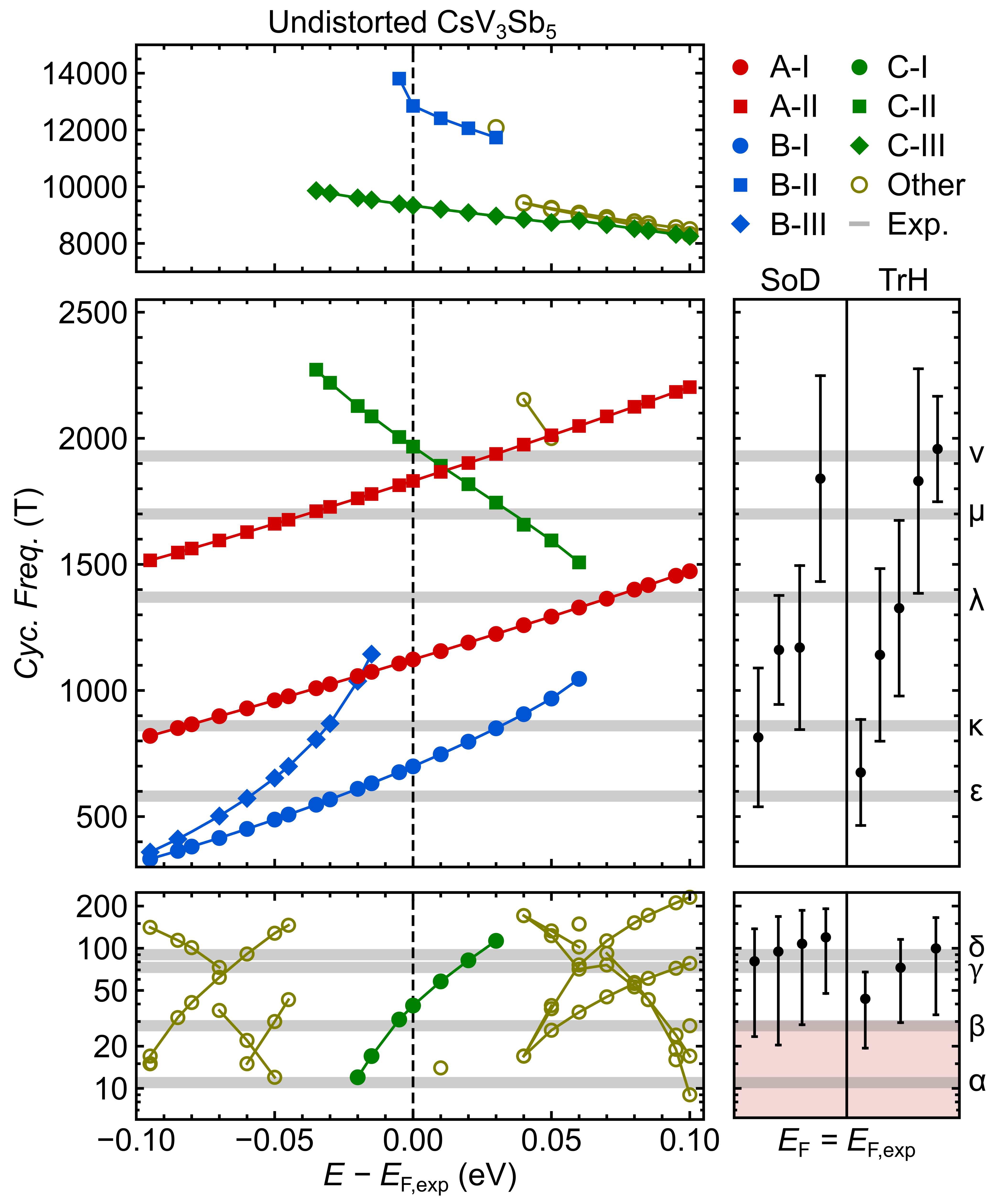}
\caption{Frequencies extracted as a function of Fermi level in undistorted CsV$_3$Sb$_5$ allow us to directly compare the experimental SdH data (grey bars) with the DFT models. The undistorted oscillation frequencies are shown as a function of Fermi level (left). Colored frequencies have been grouped by the pocket and orbit of origin. A conservative numerical estimate for the supercell orbits for the SoD and TrH structures are shown (left) at $E_\text{F}= E_\text{F,exp}$. While 5 mid-frequency orbits are indeed expected in the range $E_\text{F}\pm0.1$\,eV, no singular choice of $E_\text{F}$ produces complete agreement with experiment. However, the TrH distortion introduces a new high-frequency V $d$ orbit, similar to B-III, which provides a consistent hypothesis. The DFT resolution limit (pink shaded) indicates the frequency range below which we cannot accurately resolve closed orbits.}
\label{fig:Frequencies}
\end{figure}

Figure \ref{fig:Frequencies} summarizes all the numerical data characterizing orbits in the undistorted and distorted SoD and TrH Fermi surfaces and overlays these with the experimentally observed quantum oscillations. An exhaustive search was performed for all extremal orbits within a range of $E_\text{F}$ spanning from above and below the Fermi levels reported throughout the literature thus far ($E_\text{F,lit}$--$E_\text{F,exp}$--$E_\text{F,DFT}$). Orbits are again colored to remain consistent with their pocket designations in previous figures. For completeness, we also show orbits up to 15000\,T, though these orbits are not resolvable in the current experiments. The experimentally observed frequencies are overlaid as horizontal grey bars.

First examining comparisons of the models in the high-frequency regime ($200<f<3000$\,T), there are a total of four possible (symmetry unique) orbits at $E_\text{F,exp}$. These are the same four shown in the schematic representation. There is one orbit (B-III) which appears at slightly lower $E_\text{F}$, though it is gapped by $E_\text{F,exp}$. Comparison at $E_\text{F} = E_\text{F,exp}$, shows that, while precise quantitative agreement between all the frequencies varies between orbits, TrH layers are the only \textit{qualitative} match to 5 frequencies in this regime.  In contrast, due to the loss of the C-II orbit, the SoD distortion only recovers 4 of the 5 modes.  More detailed comparisons for $E_\text{F}$ shifted away from $E_\text{F,exp}$ will be discussed in the Discussion section.

Next, examining the low-frequency ($f<200$\,T) regime, the data show at least 4 well-defined frequencies about $E_\text{F,exp}$. However, at $E_\text{F,exp}$ there is only 1 orbit in the undistorted structure. This orbit (C-I) is predominantly comprised of vanadium $d$-orbitals associated with the Dirac-like crossings along the $\Gamma$-$K$ line. However, as shown throughout this manuscript, \textit{substantial} reconstruction is expected about the $M$ point. Characterization of the well-defined, closed orbits in the supercell models is shown to the right of the low-frequency (undistorted) structure panel. There are 4 orbits in the SoD model, and 3 in the TrH model. However, it is important to note that our unfolded supercell calculations are unable to resolve orbits with frequencies $<30$\,T.

\section{Discussion}

While precise numerical agreement between quantum oscillation frequencies and DFT-derived oscillation frequencies at $E_\text{F,exp}$ is lacking in all single layer structures modeled, the presence of the TrH structure at $E_\text{F,exp}$ within the unit cell is the only means of capturing the multiplicity of orbits within the experimentally-accessible frequency windows. This is confirmed within x-ray diffraction data which identify TrH distorted planes within the structure; however, the modulation between TrH and SoD distorted planes along the $c$-axis likely renders a more complex convolution of the orbits calculated within single layer models. Further work computationally modeling the much larger $2 \times 2 \times 4$ supercell with full spin-orbit coupling is required to definitely assess the impact of this modulated supercell and generate a more quantitative comparison to the experimentally observed orbital frequencies. 

Irrespective of the out-of-plane modulation, when combined with DFT models, our quantum oscillation data demonstrate that the reconstructed electronic states near the Fermi level are intimately tied to the vanadium $d$-orbitals. This is particularly true for the $M$ points, which are relevant for their contributions to the topologically protected surface states. 
Specifically, the observation of multiple low frequency orbits provides direct evidence of a CDW-derived reconstruction of vanadium bands endemic to the underlying Kagome lattice in CsV$_3$Sb$_5$. This finding agrees with recent ARPES results which identify gapping around the states at the M-points \cite{wang2021distinctive}.

When matching ARPES or other Fermi-surface sensitive probes, it is worthwhile to consider the impact of choosing alternative Fermi levels in the single-layer DFT models (i.e. away from $E_\text{F,exp}$), in the comparison between models and the data. In the low frequency regime for the undistorted structure, multiple orbits appear when moving both above and below $E_\text{F,exp}$, mimicking the multiple modes found in the experiment; however, this scenario can be precluded with the following arguments: (1) For the case where $E_\text{F}$ lies below $E_\text{F,exp}$---in the regime where multiple low frequency orbits appear---the multiplicity of the high-frequency modes does not match the data, as the C-II mode is absent. (2) For the case where $E_\text{F}$ lies above $E_\text{F,exp}$, the C-II and B-I orbits are quickly gapped out in the distorted structures, leaving no explanation for the five mid-frequency oscillations.  

Superlattice reflections with a propagation wave vector \textbf{q}~$=(0.5, 0.5, 0.25)$ in x-ray scattering data indicating a 2$\times$2$\times$4 superstructure with a correlation length matching the native crystallinity of the sample.  Primary Bragg reflections in the undistorted state are anisotropic due to $c$-axis broadening, and the superlattice reflections show the same degree of anisotropy.  This indicates a minimum correlation length of $\approx200$\AA\, for the out-of-plane superlattice modulation, which is born from the poorer interplane crystallinity. While the in-plane wave vector $(h, k)=(0.5, 0.5)$ matches the 3\textbf{Q} structure observed in STM, the out-of-plane component of \textbf{q} implies a four unit cell phasing along the $c$-axis. The average structure refined in $P\bar{3}$ suggests a modulation of distortion types along the c-axis. While there are a number of possible stacking sequences of SoD and TrH structures, the solution presented here almost falls naturally out of charge flipping in $P\bar{1}$ and is further sharpened by refining the structure within $P\bar{3}$.  Future work resolving the presence of orthorhombic twins in the bulk is necessary to justify pursuing lower symmetry structures, or more complex combinations of motifs (e.g. phased offsets between layers).   

The $2\times2\times4$ unit cell resolved in CsV$_3$Sb$_5$ seemingly contrasts the $2\times2\times2$ cell identified in KV$_3$Sb$_5$ \cite{yuxiaoKVS}.  Future diffraction studies will be required to fully explore this apparent difference; however one potential reason is the poorer $c$-axis crystallinity of the KV$_3$Sb$_5$ crystal explored in the earlier study.  This broadening along L can potentially mask $q_L=0.25$-type reflections or the enhanced disorder can modify the structural ground state.  While this paper was in review, another manuscript appeared by Li et al. \cite{li2021observation} instead reporting a 2$\times$2$\times$2 superstructure in CsV$_3$Sb$_5$.  We note here that our data are in agreement in the momentum space regions reported in that work. In these regions the $q_L=0.25$-type superlattice peaks are weak and below our experimental resolution. Larger surveys of reciprocal space reveal are required to map the 1/4-type $c$-axis superlattice reflections.  Disorder within a crystal can also disrupt the longer wavelength stacking, and random stacking faults are unable to create a smaller $q$ periodicity.

\section{Conclusion}

Combined DFT modeling, high-resolution x-ray scattering, and quantum oscillation measurements demonstrate that the CDW state in CsV$_3$Sb$_5$ derives from the reconstruction of the kagome-plane vanadium orbitals with an accompanying out-of-plane modulation of the distorted structure. The in-plane component of the resulting $2\times2\times4$ superstructure is best modeled using the kagome ``breathing mode”, with the SoD and TrH patterns emerging as energetically favorable structures. X-ray diffraction data are best fit via a model of modulated SoD and TrH distortions along the $c$-axis of the average structure. Quantum oscillation measurements provide a bulk probe of the electronic structure that demonstrates the CDW's reconstruction of the Fermi surface. They show the dominant role of vanadium orbitals within the kagome planes in CsV$_3$Sb$_5$ in the CDW, and support theoretical approaches drawn from minimal models focused on the kagome substructure in AV$_3$Sb$_5$ superconductors.

\section{Note added}
A work by Chen et al. \cite{CsV3SB5hall} focusing on the anomalous Hall effect in CsV$_3$Sb$_5$ also appeared during the submission of this work. They report the Fourier spectrum of the low-frequency orbits in quantum oscillation data, but do not observe the high-frequency orbits reported in our data. This could be a consequence of crystal quality, or smoothing/aliasing considerations during data collection.

\section{Acknowledgments}
Stephen Wilson gratefully acknowledges discussions with Leon Balents, Binghai Yan, and Ziqiang Wang. We gratefully thank the contributions of Matthew Benning and Michael Ruf of Bruker Corporation for their assistance in the integration of synchrotron data with Apex3 and their crystallography discussions. This work was supported by the National Science Foundation (NSF) through Enabling Quantum Leap: Convergent Accelerated Discovery Foundries for Quantum Materials Science, Engineering and Information (Q-AMASE-i): Quantum Foundry at UC Santa Barbara (DMR-1906325). The research made use of the shared facilities of the NSF Materials Research Science and Engineering Center at UC Santa Barbara (DMR- 1720256), along with the Center for Scientific Computing (NSF CNS-1725797). The UC Santa Barbara MRSEC is a member of the Materials Research Facilities Network. (www.mrfn.org). B.R.O. and P.M.S. also acknowledge support from the California NanoSystems Institute through the Elings Fellowship program. S.M.L.T has been supported by the National Science Foundation Graduate Research Fellowship Program under Grant No. DGE-1650114.  This work is based upon research conducted at the Center for High Energy X-ray Sciences (CHEXS) which is supported by the National Science Foundation under award DMR-1829070.

\bibliography{CsV3Sb5_SdH}

\end{document}